\documentclass{amsart}
\usepackage{graphicx}
\newtheorem{thm}{Theorem}[section]
\newtheorem{cor}[thm]{Corollary}
\newtheorem{lem}[thm]{Lemma}
\newtheorem{prop}[thm]{Proposition}
\theoremstyle{definition}
\newtheorem{defn}[thm]{Definition}
\theoremstyle{remark}
\newtheorem{rem}[thm]{Remark}
\numberwithin{equation}{section}
\newcommand{\Real}{\mathbb R}

\newcommand{\To}{\longrightarrow}

\begin{document}

\title[Stationary solutions of the Schr\"{o}dinger-Newton model ...]
{Stationary solutions of the Schr\"{o}dinger-Newton model - An ODE approach}%
\author{Philippe Choquard*, Joachim Stubbe** and Marc Vuffray**}%
\address{*EPFL, ITP-GE-FSB, Station 8, CH-1015 Lausanne, Switzerland}
\address{**EPFL, IMB-FSB, Station 8, CH-1015 Lausanne, Switzerland}
\email{Philippe.Choquard@epfl.ch, Joachim.Stubbe@epfl.ch,Marc.Vuffray@epfl.ch}%

\thanks{We wish to thank the referee for valuable suggestions which made the proof of our main result more efficient.}%

\date{14th February 2008}%
%\dedicatory{}%
%\commby{}%
% ----------------------------------------------------------------
\begin{abstract}
We prove the existence and uniqueness of stationary spherically
symmetric positive solutions for the Schr\"{o}dinger-Newton model
in any space dimension $d$. Our result is based on an analysis of
the corresponding system of second order differential equations.
It turns out that $d=6$ is critical for the existence of finite
energy solutions and the equations for positive spherically
symmetric solutions reduce to a Lane-Emden equation for all $d\geq
6$. Our result implies in particular the existence of stationary
solutions for two-dimensional self-gravitating particles and
closes the gap between the variational proofs in $d=1$ and $d=3$.
\end{abstract}
\maketitle
% ----------------------------------------------------------------
\section{Introduction}
We consider the Schr\"{o}dinger-Newton equations
\begin{equation}\label{SN}
    i\psi_t+\Delta\;\psi-\gamma V\psi=0,\quad \Delta\;V=|\psi|^2
\end{equation}
on $\mathbb{R}^d$ which is equivalent to the nonlinear
Schr\"{o}dinger equation
\begin{equation}\label{NLS}
    i\psi_t+\Delta \psi+\gamma(G_d(|x|)*|\psi|^2)u=0
\end{equation}
where $G_d(|x|)$ denotes the Green's function of the Laplacian on
$\Real^d$. Of physical interest are solutions having finite energy
$E$ and particle number (or charge) $N$ given by
\begin{equation}\label{energy}
E(\psi)=\frac1{2}\int_{\Real^d}|\nabla{\psi}(x,t)|^2\;dx-\frac{\gamma}{4}\iint_{\Real^{2d}}G_d(|x-y|)|\psi(x,t)|^2|{\psi}(y,t)|^2\;dxdy
\end{equation}
and
\begin{equation}\label{charge}
    N(\psi)=\int_{\Real^d}|\psi(x,t)|^2\;dx,
\end{equation}
respectively.

In the present work we study in the attractive case $\gamma >0$
the existence and uniqueness of spherically symmetric quasi
stationary solutions of the form
\begin{equation}\label{Bound states}
    \psi(t,x)=u_{\omega}(|x|)e^{-i\omega t},\quad
    u_{\omega}(|x|)>0,\quad\underset{|x|\to\infty}{\lim}u_{\omega}(|x|)=0,
\end{equation}
which we call ground states. For solutions of the form
\eqref{Bound states} we have $V(t,x)=v_{\omega}(|x|)$ and
$u_{\omega}(r),v_{\omega}(r)$ satisfy the following system of
ordinary differential equations:
\begin{equation}\label{ODE-d}
\begin{split}
    &u_{\omega}''+\frac{d-1}{r}\;u_{\omega}'=(\gamma v_{\omega}-\omega)u_{\omega}\\
    &v_{\omega}''+\frac{d-1}{r}\;v_{\omega}'=u_{\omega}^2,\quad r\geq 0.\\
    \end{split}
\end{equation}
We suppose that $u_{\omega}(0), v_{\omega}(0)$ are finite and
$u'_{\omega}(0)=v'_{\omega}(0)=0$. The latter equation implies
that $v'_{\omega}\geq 0$ and therefore for solutions $u_{\omega}$
vanishing at infinity we have $\omega-\gamma v_{\omega}(0)>0$. By
rescaling $u(r)=Au_{\omega}(r/\sigma)$,
$V(r)=B(v_{\omega}(r/\sigma)-\omega/\gamma)+1$ with
\begin{equation*}
    \sigma^2=\omega-\gamma
    v_{\omega}(0),\;A=\frac{\sqrt{\gamma}}{\sigma^2},\;
    B=\frac{\gamma}{\sigma^2}
\end{equation*}
we obtain the universal equations

\begin{equation}\label{ODE-d-re}
\begin{split}
    &u''+\frac{d-1}{r}\;u'=(V-1)u\\
    &V''+\frac{d-1}{r}\;V'=u^2\\
    \end{split}
\end{equation}
subject to the initial conditions
\begin{equation}\label{ini}
    u(0)=u_0\in\Real^+,\;u'(0)=0,\;V(0)=0, V'(0)=0.
\end{equation}

Here $d>0$ may be regarded as a continuous parameter. By analyzing
the solutions of the above initial value problem we shall prove
the following result about the existence and uniqueness of ground
states:
\begin{thm} \label{main-theorem} For any $d>0$ the system \eqref{ODE-d-re} subject to the initial conditions \eqref{ini} admits a
unique solution $(u,V)$ such that $u(r)>0$, $u'(r)<0$ on
$]0,\infty[$ and
\begin{equation}\label{decay}
     \underset{r\to\infty}{\lim}u(r)=0.
\end{equation}
If $d=6$, then $u(r)$ is explicitly given by
\begin{equation}\label{d=6}
    u(r)=\bigg(1+\frac{r^2}{24}\bigg)^{-2}.
\end{equation}
In addition, for all $d\geq 6$ the relation $u(r)=1-V(r)$ holds
and the function $u(r)$ solves the Lane-Emden equation
\begin{equation}\label{Lane-Emden}
     u''+\frac{d-1}{r}\;u'=-u^2
\end{equation}
subject to the initial conditions $u(0)=1,\;u'(0)=0$.
\end{thm}
The decay properties of these solutions will imply that they have
finite energy and particle number if and only if $d\leq 6$.

% ----------------------------------------------------------------
In the physical and mathematical literature the
Schr\"{o}dinger-Newton system in three space dimensions has a long
standing history. With $\gamma$ designating appropriate positive
coupling constants it appeared first in 1954, then in 1976 and
lastly in 1996 for describing the quantum mechanics of a Polaron
at rest by S. J. Pekar ~\cite{P54}, of an electron trapped in its
own hole by the first author ~\cite{L77} and of self-gravitating
matter by R. Penrose ~\cite{P96}. In 1977, E.Lieb ~\cite{L77}
showed the existence of a unique ground state of the form
\eqref{Bound states} in three space dimensions  by solving an
appropriate minimization problem. This ground state solution
$u_{\omega}(x), \omega>0$ is a positive spherically symmetric
strictly decreasing function. In ~\cite{L80}, P.L. Lions proved
the existence of infinitely many distinct spherically symmetric
solutions and claimed a proof for the existence of anisotropic
bound states in ~\cite{L86}.

While Lieb's existence proof can be easily extended to dimensions
$d=4$ and $d=5$, the situation is unclear for lower dimensions due
to the lack of positivity of the Coulomb interaction energy term.
For the one-dimensional problem this difficulty has been overcome
recently in ~\cite{CS2007} and the existence of a unique ground
state of the form \eqref{Bound states} has been shown by solving a
minimization problem. The two-dimensional problem, however,
remained open and so far only numerical studies are available
indicating the existence of bound states, see e.g.
~\cite{HMT2003}. Our main result proves the existence of such
solutions.

From the variational point of view the critical dimension $d=6$ is
related to a Hardy-Littlewood-Sobolev inequality of the form

\begin{equation}\label{HLS-inequality}
\iint_{\Real^{12}}|x-y|^{-4}|u(x)|^2|u(y)|^2\;dxdy\leq C
\bigg(\int_{\Real^6}|\nabla{u}(x)|^2\;dx\bigg)^2
\end{equation}
for a positive constant $C$. Our solution \eqref{d=6} is indeed an
optimizer in this inequality. Instead of proving this inequality
directly we deduce it together with the optimal constant by simply
combining two inequalities of ~\cite{L83} (see Appendix). From the
ODE point of view the system of ordinary differential equations
has a conformal invariance for $d=6$ which leads to a
one-dimensional autonomous system with a Yukawa-type interaction.
In addition, the problem of finding a positive solution can be
reduced to solving a (conformally invariant) Lane-Emden equation
~\cite{L1869},~\cite{E07}, ~\cite{C39}. We do not make use of this
property but we believe that this observation may be useful for
further studies of these equations and we give the corresponding
autonomous system in the Appendix.

To prove the main result we use a shooting method which various
authors have successfully applied to existence and uniqueness of
solutions in boundary value problems for second order nonlinear
differential equations
~\cite{BLP81},~\cite{PS83},~\cite{AP86},~\cite{K89} . Our paper is
organized as follows: In Section 2 we employ a shooting method to
prove the existence of ground states (theorem \ref{gs-existence}).
In Section 3 study their decay properties to prove uniqueness by
analyzing the Wronskian of solutions (theorem
\ref{uniqueness-in-G}). Finally, in Section 4 we prove the final
part of our main theorem including the explicit solution for $d=6$
(theorem \ref{u0=1}).

% ----------------------------------------------------------------
\section{Existence of ground states}
We begin our study with the discussion of some general properties
of solutions of \eqref{ODE-d-re} with initial values \eqref{ini}.
Standard results will guarantee local existence and uniqueness of
solutions, their continuous dependance on the initial values as
well as on the parameter $d$ and their regularity. As a
consequence of local existence and uniqueness solutions cannot
have double zeros. We shall frequently apply these properties in
the sequel as well as the following integral equations for $u'$
and $V'$:

\begin{equation}\label{int-eq}
\begin{split}
    &u'(r)=\frac{1}{r^{d-1}}\int_{0}^r(V(s)-1)u(s)s^{d-1}\;ds\\
    &V'(r)=\frac{1}{r^{d-1}}\int_0^ru^2(s)s^{d-1}\;ds.\\
    \end{split}
\end{equation}
Viewed as a mechanical system we can associate an energy to
\eqref{ODE-d-re} given by
\begin{equation}\label{energy1}
    2\mathcal{E}=u'^2+u^2+\frac1{2}V'^2-Vu^2
\end{equation}
which satisfies
\begin{equation*}
    \mathcal{E}'=-\frac{d-1}{r}\;u'^2-\frac{d-1}{2r}\;V'^2.
\end{equation*}
Therefore $\mathcal{E}$ is a constant of motion if $d=1$. However,
we shall not use this property in the present work.

For the initial condition $u_0>0$ of the solution $(u,V)$ we
consider the following mutually disjoint sets:
\begin{defn}
\begin{equation}\label{N-set}
    \mathcal{N}=\{u_0\in\Real_+: \exists r_0>0\;\text{such that}\; u(r_0)<0 \;\text{and}\; u'(r)<0 \;\text{on}
    \;]\,0,r_0]\;\},
\end{equation}
\begin{equation}\label{G-set}
    \mathcal{G}=\{u_0\in\Real_+: u\geq
    0,\;\underset{r\To\infty}{\lim}u(r)=0\},
\end{equation}
\begin{equation}\label{P-set}
    \mathcal{P}=\{u_0\in\Real_+: \exists r_1>0\;\text{such that}\; u'(r_1)>0 \;\text{and}\; u(r)>0 \;\text{on}
    \;]\,0,r_1]\;\}.
\end{equation}
\end{defn}
In order to see that $\mathcal{G}$ and $\mathcal{P}$ are disjoint
note that since $u''(0)=-du_0$ all solutions start strictly
decreasing. Therefore any solution with initial condition $u_0$ in
$\mathcal{P}$ has a local minimum before $r_1$ where $V\geq 1$.
From \eqref{int-eq} we deduce that $V$ is strictly increasing.
Therefore $u'$ will remain positive and bounded away from zero
after $r_1$ and $u$ becomes unbounded. If $u_0\notin
\mathcal{N}\cup\mathcal{P}$, then $u\geq 0$ and $u'\leq 0$.
Therefore it has a limit as $r$ tends to infinity which must be
zero. This implies $\mathcal{N}\cup
\mathcal{G}\cup\mathcal{P}=\Real_+$. From the continuous
dependance on initial values we deduce that
$\mathcal{N},\mathcal{P}$ are open sets.

Our main result theorem \ref{main-theorem} states that
$\mathcal{G}$ consists of exactly one element. Obviously,
$\mathcal{G}$ is nonempty if $\mathcal{N}$ and $\mathcal{P}$ are
nonempty which we show in the following lemma.
\begin{lem}\label{PN-lemma} For any $d>0$ the sets $\mathcal{N}$ and
$\mathcal{P}$ are nonempty. In particular, $]0,1[\subset
\mathcal{N}$.
\end{lem}
\begin{proof} We consider the function
\begin{equation}\label{u+V-1}
    \phi=u+V-1.
\end{equation}
It satisfies the differential equation
\begin{equation*}
    \phi''+\frac{d-1}{r}\;\phi'=u\phi
\end{equation*}
and admits the Taylor expansion
\begin{equation*}
    \phi(r)=(u_0-1)\big(1+\frac{u_0r^2}{2d}\big)+o(r^2).
\end{equation*}
Let $u_0<1$ and suppose $u_0\notin \mathcal{N}$. Then $\phi$ is
negative and strictly decreasing for all $r>0$ sufficiently small.
By hypothesis $u$ remains strictly positive. Consequently, $\phi$
cannot have a critical point since then $\phi''=u\phi<0$ which is
impossible. We conclude that $\phi(r)<\phi(0)=u_0-1$ for all $r>0$
or equivalently, $u(r)+V(r)<u_0$. Hence $u_0\in \mathcal{G}$ and
$V(r)< u_0$ for all $r>0$. Since $V$ is always strictly increasing
$V_{\infty}:=\underset{r\To\infty}{\lim}V(r)$ exists and
$V_{\infty}\leq u_0<1$. We consider the function
$z:=-\frac{u'}{u}$. Since $u_{0}\in\mathcal{G}$ it follows that
$z$ exists for all $r>0$ and $z(r)>0$ for all $r>0$. It satisfies
the differential equation
\begin{equation*}
    z'=z^2-\frac{d-1}{r}\;z+1-V.
\end{equation*}
Choose $\tilde{r}$ such that $\frac{d-1}{r}\leq
\sqrt{2(1-V_{\infty})}$ for all $r\geq \tilde{r}$. Then for $r\geq
\tilde{r}$ we have
\begin{equation*}
    z'\geq
    \frac1{2}z^2+\Big(\frac1{2}z^2-\sqrt{2(1-V_{\infty})}z+1-V_{\infty}\Big)\geq\frac1{2}z^2.
\end{equation*}
This implies that $z$ blows up in finite time which is impossible.
Hence $u_0\in\mathcal{N}$.

Next we want to show that $u_0\in \mathcal{P}$ for $u_0$
sufficiently large. Suppose on the contrary that $\mathcal{P}$ is
empty and let $u_0>1$. Denote $]0,R_0[$ the maximal interval where
$u>0$ and $u'<0$. Therefore from equation \eqref{int-eq} for $V'$
we obtain the bounds
\begin{equation*}
   \frac{u(r)^2r}{d}\leq V'(r)\leq \frac{u_0^2r}{d}\quad\text{on}\;]0,R_0[.
\end{equation*}
Integrating these inequality and using again that $u$ is
decreasing yields the following estimates for $V$:
\begin{equation*}
   \frac{u(r)^2r^2}{2d} \leq V(r)\leq \frac{u_0^2r^2}{2d}\quad\text{on}\;]0,R_0[.
\end{equation*}
By a similar reasoning as before we see that the function $\phi$
defined in \eqref{u+V-1} is strictly increasing on $]0,R_0[$.
Hence
\begin{equation*}
    u(r)>u_0-V(r)\quad\text{on}\;]0,R_0[.
\end{equation*}
Let $r_0=\sqrt{2d/u_0}$. Inserting the upper bound for $V$ we get
\begin{equation*}
    u(r)>u_0\big(1-\frac{r^2}{r_0^2}\big)\quad\text{on}\;]0,r_0[
\end{equation*}
and $r_0\leq R_0$. We want to show that $u'(r_0)>0$ provided $u_0$
is sufficiently large which yields the desired contradiction.
Using our bounds on $u$ and $V$ in \eqref{int-eq} we obtain
\begin{equation*}
    \begin{split}
u'(r_0)&=\frac{1}{r_0^{d-1}}\int_{0}^{r_0}(V(r)-1)u(r)r^{d-1}\;dr\\
&\geq \frac{1}{2d\,r_0^{d-1}}\int_{0}^{r_0}u^3(r)r^{d+1}\;dr-\frac{u_0r_0}{d}\\
&\geq \frac{u_0^3}{2d\,r_0^{d-1}}\int_{0}^{r_0}\big(1-\frac{r^2}{r_0^2}\big)^3r^{d+1}\;dr-\frac{u_0r_0}{d}\\
&\geq \frac{u_0r_0}{d}\bigg(du_0\int_{0}^{1}(1-s^2)^3s^{d+1}\;ds-1\bigg).\\
    \end{split}
\end{equation*}
We conclude that $u'(r_0)>0$ for $u_0$ sufficiently large which
contradicts the assumption that $\mathcal{P}$ is empty.
\end{proof}
Hence we have proved by the preceding lemma the existence of
ground states:
\begin{thm}\label{gs-existence}
For any $d\geq 1$, the set $\mathcal{G}$ is nonempty, that is
there is a solution $(u,V)$ \eqref{ODE-d-re} subject to the
initial conditions \eqref{ini} such that $u(r)>0$, $u'(r)<0$ on
$]0,\infty[$ and $\underset{r\to\infty}{\lim}u(r)=0$.
\end{thm}
\begin{proof}
It remains to prove $u(r)>0$, $u'(r)<0$. The first property
follows from the fact there are no double zeros. If $u$ has a
first critical point $r_1>0$, then $V(r_1)\geq 1$ and since $V$ is
strictly increasing (see \eqref{int-eq}) it follows again from
\eqref{int-eq} that $u(r)>0$ for all $r>r_1$ which is impossible.
Hence $u'<0$.
\end{proof}

\section{Uniqueness of ground states}
In this section we prove that $\mathcal{G}$ has exactly one
element. First of all, we show that if $\mathcal{G}$ had more than
one element the corresponding solutions cannot cross. This is an
immediate consequence of the following lemma which states that any
two solutions of the initial value problem
\eqref{ODE-d-re},\eqref{ini} cannot cross as long as they stay
positive.
\begin{lem}\label{no-cross} Let $u_{2}(0)>u_{1}(0)>0$ and suppose that $u_{2}(r),u_{1}(r)$ exist on $[0,R]$ such that $u_{1}(r)\geq 0$ on $[0,R]$.
Then $u_{2}(r)>u_{1}(r)$ for all $r\in [0,R]$.
\end{lem}
\begin{proof} We consider the Wronskian of $u_1,u_2$ defined by
\begin{equation}\label{Wronskian}
    w(r)=u_2'(r)u_1(r)-u_1'(r)u_2(r).
\end{equation}
Then $w$ satisfies the differential equation
\begin{equation}\label{wronskian-ode}
   w'+\frac{d-1}{r}\;w=(V_2-V_1)u_1u_2.
\end{equation}

Suppose there is $\bar{r}\in [0,R]$ such that $u_2(r)>u_1(r)$ on
$[0,\bar{r}[$ and $u_1(\bar{r})=u_2(\bar{r})\geq 0$. Then
\begin{equation*}
    w(\bar{r})=(u_2'(\bar{r})-u_1'(\bar{r}))u_1(\bar{r})\leq 0.
\end{equation*}
On the other hand we have
\begin{equation*}
     V_2'(r)-V_1'(r)=\frac{1}{r^{d-1}}\int_0^r(u_2^2(s)-u_1^2(s))s^{d-1}\;ds>0
\end{equation*}
on $]0,\bar{r}]$ and therefore  $V_2(r)>V_1(r)$ on $]0,\bar{r}]$.
We conclude then from the differential equation
\eqref{wronskian-ode} for $w$ that $w r^{d-1}$ is strictly
increasing on $]0,\bar{r}]$ and since $w(0)=0$ we must have
$w(\bar{r})>0$ which is the desired contradiction.
\end{proof}
\begin{rem}
From the no-crossing property stated in lemma \ref{no-cross} it
follows immediately that $\mathcal{N}, \mathcal{P}$ are intervals.
More precisely, $\mathcal{N}=]0,a[, \mathcal{P}=]b,\infty[$ with
$0\leq a\leq b\leq \infty[$. Uniqueness of ground states is then
equivalent to $a=b$.
\end{rem}
The important conclusion from lemma \ref{no-cross} is that two
different ground state solutions cannot intersect. From the
differential equation \eqref{wronskian-ode} for their Wronskian
$w(r)$ we see that $w(r)r^{d-1}$ is a nonnegative strictly
increasing function. However, we shall prove in the sequel that
$w(r)r^{d-1}$ vanishes at infinity which yields the desired
contradiction. Therefore we have to analyze the decay properties
of ground states at infinity.

Since $V$ is always strictly increasing
$V_{\infty}:=\underset{r\To\infty}{\lim}V(r)$ exists (including
the case $V_{\infty}=+\infty$) and $V(r)<V_{\infty}$ for all
$r>0$.

\begin{lem}\label{u-V-lemma} Let $u_0\in\mathcal{G}$. Then $1\leq u_0\leq
V_{\infty}$. In particular, $u_0=1$ if and only if $V_{\infty}=1$
and in this case $u=1-V$.
\end{lem}
\begin{proof}
Since $\mathcal{N}\subset ]0,1[$ by lemma \ref{PN-lemma} the
inequality $1\leq u_0$ is obvious. To prove the second inequality
it is sufficient to consider the case $V_{\infty}<+\infty$. We
consider the function $\xi$ defined by
\begin{equation*}
    \xi=u+V-V_{\infty}.
\end{equation*}
Obviously, $\xi(0)=u_0-V_{\infty}$, $\xi'(0)=0$ and
$\underset{r\to\infty}{\lim}\xi(r)=0$.  The function $\xi$
satisfies the differential equation
\begin{equation*}
    \xi''+\frac{d-1}{r}\;\xi'=u\xi+(V_{\infty}-1)u.
\end{equation*}
and admits the Taylor expansion
\begin{equation*}
    \xi(r)=u_0-V_{\infty}+\frac{u_0(u_0-1)r^2}{2d}+o(r^2).
\end{equation*}
Suppose $u_0>V_{\infty}$. Then $\xi(r)$ is positive and strictly
    increasing for $r>0$ sufficiently small. Therefore $\xi$ must
    have a critical point $r_1$ with $\xi(r_1)>u_0-V_{\infty}$ and $\xi''(r_1)\leq
    0$ which is impossible since $u>0$.

For the initial condition $u(0)=1$ we have thanks to the
uniqueness of solutions for the initial value problem
\eqref{ODE-d-re}, \eqref{ini} that $u=1-V$.
\end{proof}

In the following lemma we determine the asymptotic behavior of
ground states.

\begin{lem} \label{G-riccati-decay} Let $u_{0}\in\mathcal{G}$. Then

\begin{equation*}
    \underset{r\To\infty}{\lim}\frac{u'}{u}=-\sqrt{V_{\infty}-1}.
\end{equation*}
Moreover, if $1<V_{\infty}\leq\infty$, then for any $\kappa\in
]\;0,\sqrt{V_{\infty}-1}\,[\;$,
\begin{equation*}
   \underset{r\To\infty}{\lim\sup}\;u(r)e^{\kappa r}<\infty.
\end{equation*}
\end{lem}
\begin{proof}
First off all, let $V_{\infty}<\infty$. We consider the function
$z:=-\frac{u'}{u}$ which is well defined for all $r\geq 0$ and
satisfies the differential equation
\begin{equation*}
    z'=z^2-\frac{d-1}{r}\;z+1-V.
\end{equation*}
Now choose $\tilde{r}$ such that $\frac{d-1}{r}\leq
\frac1{2}\sqrt{V_{\infty}}$ for all $r\geq \tilde{r}$. Consider
the direction field in the $(r,z)$ plane for the preceding
differential equation. In the set $r\geq \tilde{r}$, $z\geq
2\sqrt{V_{\infty}}$ we have
\begin{equation*}
    z'\geq
    \frac1{2}z^2+\Big(\frac1{2}z^2-\frac1{2}\sqrt{V_{\infty}}z+1-V_{\infty}\Big)\geq\frac1{2}z^2+1.
\end{equation*}
It follows that, should $z(r)$ ever enter this region, it would
blow up at finite time after $\tilde{r}$ which is impossible.
Hence $z$ remains bounded. This also implies

\begin{equation*}
    \underset{r\To\infty}{\lim}u'(r)=0.
\end{equation*}

Therefore we may apply l'H\^{o}spital's rule. We obtain
\begin{equation*}
    \underset{r\To\infty}{\lim}z^2=\underset{r\To\infty}{\lim}\frac{u''}{u}=\underset{r\To\infty}{\lim}\Big(\frac{d-1}{r}\;z+V-1\Big)=V_{\infty}-1.
\end{equation*}
Finally, if $V_{\infty}$ is infinite, then $z$ is also unbounded
since otherwise applying l'H\^{o}spital's rule as above yields the
desired contradiction. This proves the first part of the lemma.

Now let $V_{\infty}>1$. Then for any $\kappa\in
]\;0,\sqrt{V_{\infty}-1}\;[$ and $r$ sufficiently large,
$-\frac{u'}{u}\geq \kappa$ and the proof is completed by
integrating this inequality and taking exponentials on both sides.
\end{proof}
\begin{rem} If $V_{\infty}=+\infty$ the asymptotic behavior of
ground states can be given more precisely. Indeed, by analyzing
the differential equation for $Z:=-\frac{u'}{u}V^{-1/2}$

and taking into account that $\underset{r\To\infty}{\lim}V'/V=0$
it can be easily shown by mimicking the proof of the preceding
lemma that $\underset{r\To\infty}{\lim}Z=1$.
\end{rem}
Now we are in position to prove our uniqueness result:

\begin{thm} \label{uniqueness-in-G}The set $\mathcal{G}$ has exactly one element.
\end{thm}
\begin{proof}
Let $u_{1}(0),u_{2}(0)\in\mathcal{G}$ such that
$u_{2}(0)>u_{1}(0)$. By lemma \ref{no-cross} the corresponding
solutions $u_1,u_2$ cannot intersect and we have $u_2(r)>u_1(r)>0$
for all $r\geq 0$. From the differential equation
\eqref{wronskian-ode} for their Wronskian $w(r)$ we see that
$w(r)r^{d-1}$ is a nonnegative strictly increasing function since
\begin{equation*}
   \Big(wr^{d-1}\Big)'=(V_2-V_1)u_1u_2r^{d-1}>0
\end{equation*}
and $w(0)=0$. On the other hand, we claim that
\begin{equation*}
   \underset{r\To\infty}{\lim}\;wr^{d-1}=0.
\end{equation*}
Indeed, by lemma \ref{u-V-lemma} we have
$\underset{r\to\infty}{\lim}V_2(r)>1$. Trivially, $V(r)\leq
\frac{u_2(0)^2r^2}{2d}$. From the integral equation \eqref{int-eq}
for $u'$,
\begin{equation*}
    u_2'(r)r^{d-1}=\int_{0}^r(V_2(s)-1)u_2(s)s^{d-1}\;ds
\end{equation*}
and the decay properties of $u_2$ given in lemma
\ref{G-riccati-decay} it follows then that $u'_2r^{d-1}$ and
$u_2r^{d-1}$ are uniformly bounded. Therefore
\begin{equation*}
  |w(r)r^{d-1}|\leq|u_1||u_2'r^{d-1}|+|u_1'||u_2r^{d-1}|\leq
  c_1|u_1|+c_2|u_1'|
\end{equation*}
for some positive constants $c_1,c_2$ which concludes the proof.
\end{proof}

%----------------------------------------------------------------

\section{Further properties of ground states }

For the initial condition $u(0)=1$ we have thanks to the
uniqueness of solutions for the initial value problem
\eqref{ODE-d-re}, \eqref{ini} that $u=1-V$ and therefore we have
to solve the following initial value problem
\begin{equation}\label{ODE-ini-reduced}
    u''+\frac{d-1}{r}\;u'=-u^2, \quad u(0)=1,\;u'(0)=0.
\end{equation}
This is the $d$-dimensional Lane-Emden equation. The behavior of
its solutions has been widely studied in the mathematical
literature. However, in the following we will give alternative
proofs of the results relevant for our work. Analyzing the
behavior of the solutions  of the initial value problem
\eqref{ODE-ini-reduced} we prove the final part of our main
result:

\begin{thm}\label{u0=1} If $d\geq 6$, then $1\in\mathcal{G}$ and if $d<6$, then $1\in\mathcal{N}$.
In particular for $d=6$ the ground state solution is explicitly
given by
\begin{equation}
    u(r)=\bigg(1+\frac{r^2}{24}\bigg)^{-2}.
\end{equation}

\end{thm}

\begin{proof}
Since $]0,1[ \subset\mathcal{N}$ by lemma \ref{PN-lemma} and
$\mathcal{P}$ is open, $1\notin\mathcal{P}$ for any $d>0$. Now let
$d\geq 6$. We introduce a new Lyapunov function $L(r)$ defined by
\begin{equation}\label{L-ode-reduced}
    L(r)=E(r)r^d+\frac{d}{3}u(r)u'(r)r^{d-1}.
\end{equation}
Then $L(0)=0$ and
\begin{equation*}
    L'(r)=-\frac{d-6}{6}u'^2(r)r^{d-1}\leq 0
\end{equation*}
since $d\geq 6$. Hence $L(r)\leq 0$ for all $r\geq 0$. We suppose
that $1\in \mathcal{N}$. Then there exists $r_0>0$ such that
$u(r_0)=0$. Computing $L$ at this point we get
\begin{equation*}
    L(r_0)=\frac1{2}u'^2(r_0)r_0^d>0
\end{equation*}
which is impossible. The explicit solution \eqref{d=6} for $d=6$
is readily verified (see also Appendix A).

Let $d<6$ and assume $1\in\mathcal{G}$. We may then use the Milne
variables (see e.g. ~\cite{M32},~\cite{C39})
\begin{equation*}
    y:=\frac{-ru'}{u},\quad z:=-\frac{-ru^2}{u'}
\end{equation*}
which are well defined for all $r\geq 0$. Indeed, $y(0)=0,z(0)=d$
and $y,z>0$ for all $r>0$. They satisfy the differential equations
\begin{equation*}
    y'=\frac{y}{r}(2-d+y+z),\quad z'=\frac{z}{r}(d-2y-z).
\end{equation*}
If $d\leq 2$, then the first differential equation implies that
$y$ blows up in finite time which is impossible. If $d>2$, then
$y$ and $z$ remain bounded, which implies that for all $r$
sufficiently large
\begin{equation*}
    u(r)\leq Cr^{-2},\quad -u'\leq Cr^{-3}
\end{equation*}
for an appropriate constant $C>0$. We consider now the Lyapunov
functional $L(r)$ defined in \eqref{L-ode-reduced}. Since $d<6$
The bounds on $u,u'$ imply that
\begin{equation*}
    \underset{r\to\infty}{\lim}L(r)=0
\end{equation*}
On the other hand, $L(0)=0$ and
\begin{equation*}
    L'(r)=-\frac{d-6}{6}u'^2(r)r^{d-1}> 0
\end{equation*}
which yields the desired contradiction.
\end{proof}

\begin{cor}
Let $d<6$ and $u_0\in\mathcal{G}$. Then $V_{\infty}>u_0>1$.
\end{cor}
\begin{proof}
The inequality $u_0>1$ is an immediate consequence of the previous
theorem. The strict inequality $V_{\infty}>u_0$ follows from the
proof of lemma \ref{u-V-lemma}: If $V_{\infty}=u_0>1$, then
$\xi=u+V-V_{\infty}$  is still positive increasing for $r>0$ which
is impossible.
\end{proof}

%----------------------------------------------------------------

%------------------------------------------------------------------------------------

%------------------------------------------------------------------------------------------

%-------------------------------------------------------------------------------------
\appendix
\section{Transformation to an autonomous system}
Putting $u(r)=e^{-2s}\phi(s)$, $V(r)-1=e^{-2s}W(s)$ with $s=\ln r$
the system \eqref{ODE-d-re} transforms into the autonomous system
\begin{equation}\label{ODE-d-transformed}
    \begin{split}
    &\ddot{\phi}+(d-6)\;\dot{\phi}-2(d-4)\phi=W\phi\\
    &\ddot{W}+(d-6)\;\dot{W}-2(d-4)W=\phi^2\\
    \end{split}
\end{equation}
subject to the boundary conditions
\begin{equation}\label{ini-transformed}
    \underset{s\to-\infty}{\lim}e^{-2s}\phi(s)=u_0\in\Real^+,\underset{s\to-\infty}{\lim}e^{-2s}W(s)=-1.
\end{equation}
For $u_0\in\mathcal{G}$ we have the asymptotic behavior
\begin{equation}\label{sol-G-transformed}
    \underset{s\to\infty}{\lim}e^{-2s}\phi(s)=0,\underset{s\to\infty}{\lim}e^{-2s}W(s)=V_{\infty}-1.
\end{equation}
 We can
associate an energy of the system given by
\begin{equation}\label{energy11}
    2E=\dot{\phi}^2-2(d-4)\phi^2+\frac1{2}\dot{W}^2-(d-4)W^2-W\phi^2
\end{equation}
which satisfies
\begin{equation}\label{energy21}
    \dot{E}=-(d-6)(\dot{\phi}^2+\frac{1}{2}\;\dot{W}^2).
\end{equation}
We should note that though the new system is invariant under
translations in $s$, the boundary conditions
\eqref{ini-transformed} break this symmetry and therefore the
solutions are not translation invariant. However, $e^{-2s}\phi(s)$
and $e^{-2s}W(s)$ are translation invariant which corresponds to
the dilation invariance of the original system \eqref{ODE-d-re}.

If $d=6$, then system \eqref{ODE-d-transformed} is Hamiltonian and
solutions satisfying the boundary conditions
\eqref{ini-transformed} have zero energy. We look for a solution
such that $\phi=-W$ (i.e. $u_0=1$). Then the zero energy condition
reads
\begin{equation}\label{solving-d=6-1}
    \dot{\phi}^2-4\phi^2+\frac{2}{3}\phi^3=0.
\end{equation}
Since $\phi$ vanishes at $\pm\infty$ there is $s_0\in\mathbb{R}$
such that $\dot{\phi}(s_0)=0$. Equation \eqref{solving-d=6-1}
yields then ${\phi}(s_0)=6$. Integrating \eqref{solving-d=6-1} we
get
\begin{equation}\label{solving-d=6-2}
    2(s-s_0)=\int_{6}^{\phi(s)}\frac{du}{u\sqrt{1-u/6}}
\end{equation}
and we obtain the solution
\begin{equation}\label{solving-d=6-3}
   \phi(s)=\frac{6}{\cosh^2(s-s_0)}.
\end{equation}
The boundary condition \eqref{ini-transformed} yields
$e^{2s_0}=24$ which after changing variables gives the solution
\eqref{d=6}.
\section{A Sobolev inequality}
In this part we make the simple observation that the
Sobolev-inequality \eqref{HLS-inequality} and its optimizers can
be obtained by combining two inequalities of ~\cite{L83}. We
denote by $H^1(\mathbb{R}^d)$ the space of functions
$f:\mathbb{R^d}\to \mathbb{C}$ for which $f$ and $\nabla f$ are
square-integrable.
\begin{prop} Let $d>4$. For all $f,g\in H^1(\mathbb{R}^d)$
\begin{equation}
    \iint_{\Real^{2d}}|x-y|^{-4}|f(x)|^2|g(y)|^2\;dxdy\leq C_d
\int_{\Real^d}|\nabla{f}(x)|^2\;dx\int_{\Real^d}|\nabla{g}(x)|^2\;dx
\end{equation}
with
\begin{equation}
   C_d=\frac{4(d-1)}{d^2(d-2)^2(d-4)}
\end{equation}
and optimizers (up to translations and dilations) given by
\begin{equation}
   f(x)=g(x)=(1+|x|^2)^{-\frac{d-2}{2}}.
\end{equation}
\end{prop}
\begin{proof}
We apply inequality (1.1) of ~\cite{L83} with $p=t=\frac{d}{d-2}$
to get
\begin{equation*}
    \iint_{\Real^{2d}}|x-y|^{-4}|f(x)|^2|g(y)|^2\;dxdy\leq
    N_{p,4,d}||f||_{2p}^2||g||_{2p}^2
\end{equation*}
and $N_{p,4,d}$ given by eq. (3.2) of ~\cite{L83} and then the
Sobolev inequality (1.2) with $K_d$ given in eq. (4.11) of
~\cite{L83}. Both inequalities have the same optimizers which
concludes the proof.
\end{proof}
% ----------------------------------------------------------------
\bibliographystyle{amsplain}
\bibliography{references}

\end{document}